\documentclass[twocolumn,showpacs,prl]{revtex4}%
\usepackage{amssymb}
\usepackage[dvips]{color,graphics,epsfig}
\usepackage{graphicx}
\usepackage{dcolumn}
\usepackage{bm}
\usepackage{amsmath}
\usepackage{amsfonts}
\usepackage{times}%
\providecommand{\U}[1]{\protect\rule{.1in}{.1in}}

\begin{document}
\title{Andreev spectra and subgap bound states in multiband superconductors}
\author{A.A. Golubov$^{1}$, A. Brinkman$^{1}$, Yukio Tanaka$^{2}$, I.I. Mazin$^{3}$, O.V.
Dolgov$^{4}$}
\affiliation{$^{1}$Faculty of Science and Technology and MESA+ Institute for Nanotechnology,
University of Twente, 7500 AE Enschede, The Netherlands}
\affiliation{$^{2}$Department of Applied Physics, Nagoya University, Nagoya, 464-8603, Japan}
\affiliation{$^{3}$Naval Research Laboratory, 4555 Overlook Ave. SW, Washington, DC 20375, USA}
\affiliation{$^{4}$Max-Planck-Institut f\"{u}r Festk\"{o}rperforschung, D-70569 Stuttgart, Germany}
\date{\today }

\begin{abstract}
A theory of Andreev conductance is formulated for junctions involving normal
metals (N) and multiband superconductors (S) and applied to the case of
superconductors with nodeless extended $s_{\pm}$-wave order parameter
symmetry, as possibly realized in the recently discovered ferropnictides. We
find qualitative differences from tunneling into $s$-wave or $d$-wave
superconductors that may help to identify such a state. First, interband
interference leads to a suppression of Andreev reflection in the case of a
highly transparent N/S interface and to a current deficit in the tunneling
regime. Second, surface bound states may appear, both at zero and at non-zero
energies. These effects do not occur in multiband superconductors without
interband sign reversal.
\end{abstract}

\maketitle

The recent discovery of high-$T_{c}$ superconductivity in ferropnictides has
been a major event in solid state physics. The first theoretically proposed
pairing symmetry for this compound, $s$-wave with sign-reversing order
parameter ($s_{\pm}$) has been followed by a number of theoretical papers
substantiating this proposal at various degrees of sophistication and
exploring ramifications of the proposed state \cite{Mazin}. Within
this proposal, two sets of Fermi surfaces are distinguished: the hole
Fermi surfaces around $\Gamma$ and the electron Fermi surfaces around M. The
$\pi$ phase difference between the superconducting condensates is thought to
be induced by spin-fluctuations. Experimental evidence has been favorable to
the $s_{\pm}$ model so far, but is still ambiguous.

Andreev spectroscopy is a strong experimental probe of the superconducting order
parameter. But in the case of the ferropnictides, both nodeless as well as
nodal superconductivity have been inferred from the absence \cite{PCAR1} or
presence \cite{PCAR2} of zero-bias conductance peaks. Also, point-contact
spectroscopy has both provided evidence for a single gap as well as multiple
gaps \cite{PCAR1,PCAR2}. One of many questions to be asked in this connection
is how possible interference between the two bands in the ferropnictides may
affect the Andreev conductance spectra. Can the interference phenomenon be used to distinguish the
$s_{\pm}$ state from other scenarios? To address these questions, a
generalized theory of Andreev conductance is needed, relevant also to other multiband systems.

Surface phenomena in $s_{\pm}$ superconductors have attracted considerable
recent attention \cite{Choi,Linder,Tsai,Feng,Ghaemi}. Certain limiting cases
were considered, but no general calculation of Andreev and tunneling current,
properly accounting for the effect of interference between the two relevant
bands, have been published so far. This we provide in the present Letter. 
The formation of bound states at a free surface of an $s_{\pm}$ superconductor, 
at an $S_{\pm}/N/S$, and at an $N/S/S_{\pm}$ junction was found in Refs.~\onlinecite{Ghaemi}, 
\onlinecite{Tsai} and \onlinecite{Feng}, respectively. However, the conditions for such a
bound state and its effect on Andreev and tunneling conductance were not
addressed in these papers. Ref. \onlinecite{Choi} found an enhancement of the density
of states at zero energy in a thin N layer on the top of an $s_{\pm}$
superconductor, but it is unclear whether this numerical result is related to
bound states or not, since \emph{finite energy} rather than \emph{zero-energy}
bound states were predicted in \cite{Ghaemi,Tsai,Feng}. Finally, Ref.
\onlinecite{Linder} considered the same problem as ours, the conductance spectra in
an $N/S_{\pm}$ junction, but did not find any new effects compared to regular
Andreev reflection. This may be related to not properly accounting for any
interference effect. At the moment, a general analytical unrestricted
treatment of an interface in an arbitrary $s_{\pm}$ superconductor seems
necessary to clarify the existing confusion.

We have studied Andreev conductance in an $N/S_{\pm}$
junction by including in the classical \textquotedblleft BTK\textquotedblright%
\ model \cite{BTK} the interference between the excitations in a two-band
superconductor at arbitrary interface transparency. Applying our extended BTK
model to the $s_{\pm}$ scenario we have found qualitatively new effects. For a
fully transparent interface, the destructive interband interference leads to a
strong suppression of Andreev reflection, in striking contrast to the
conventional case. In the tunneling regime, two new effects are found: (a) a
current deficit at high bias voltage, which is also due to destructive
interband interference, and (b) Andreev bound states (ABS) similar to those
responsible for the zero-bias anomaly (ZBA), a well known fingerprint of
$d$-wave superconductors \cite{Bound_states}. However, instead of a ZBA, a
peak of similar origin may appear (depending on the parameters) at a finite
energy, that can easily be mistaken for an extra gap.

A ballistic Andreev contact can be modeled by a one-dimensional conductor,
whose right half ($x>0$) is a two-band metal (two different states at the
Fermi level, one with the wave vector $p$ and the other with $q$), and the
left half is a simple metal. The wave function at the energy $E_{F}$ in the
left half is $\Psi_{L}(x)=\psi_{k}(x)+b\psi_{-k}(x),$ where the first term is
the incident Bloch wave and the second term the reflected Bloch wave. The wave
function on the right-hand side is $\Psi_{R}(x)=c[\phi_{p}(x)+\alpha_{0}%
\phi_{q}(x)].$ Here $p$ and $q$ are the Fermi vectors for the two bands,
$\phi$ is the Bloch function in the two-band metal, and the mixing coefficient
$\alpha_{0}$ defines the ratio of probability amplitudes for an electron
crossing the interface from the left to tunnel into the first or second band
on the right. Similar problems have been studied in the theory of the
tunneling magnetoresistance, where the leads are usually multiband $d$-metals.

The main conceptual pitfall here is that the standard approaches to tunneling
assume the wave functions to be plane waves. However, there cannot be two
different plane waves propagating in the same direction with the same energy.
This may only occur when the wave functions are $Bloch$ waves - which they are
in reality. It has been realized in the last decade that there is not a single
factor that would define the relative tunneling probability of the two Bloch waves. If the wave
vector parallel to the interface is not conserved (it usually is not, except
perhaps for epitaxially grown contacts), one factor is the number of tunneling
channels in each band, proportional to $\left\langle Nv_{\perp}\right\rangle
$, where $N$ is the density of states and $v_{\perp}$ is the Fermi velocity
component normal to the interface. Even more important is the character of
these Bloch wave functions. \textit{E.g.,} states of different parity on the
right and on the left sides of the interface hardly overlap, so that even a
weak interface barrier will strongly suppress tunneling from particular
bands. With this in mind, we keep $\alpha_{0}$ arbitrary and present the
results for different cases. One implication is that the observable tunneling
spectra may actually change drastically from contact to contact, as the
interface properties change. Indeed, there are indications that this may be
the case \cite{exps}.

At a normal metal (N) - superconductor (S) contact, in the case of
the two-gap model with unequal s-wave gaps, one can write
\begin{align}
\Psi &  =\Psi_{N}\theta(-x)+\Psi_{S}\theta(x),\nonumber\\
\Psi_{N} &  =\psi_{k}\left(
\begin{array}
[c]{c}%
1\\
0
\end{array}
\right)  +a\psi_{k}\left(
\begin{array}
[c]{c}%
0\\
1
\end{array}
\right)  +b\psi_{-k}\left(
\begin{array}
[c]{c}%
1\\
0
\end{array}
\right)  ,\\
\Psi_{S} &  =c\left[  \phi_{p}\left(
\begin{array}
[c]{c}%
u_{1}\\
v_{1}e^{-i\varphi_{1}}%
\end{array}
\right)  +\alpha_{0}\phi_{q}\left(
\begin{array}
[c]{c}%
u_{2}\\
v_{2}e^{-i\varphi_{2}}%
\end{array}
\right)  \right]  \nonumber\\
&  +d\left[  \phi_{-p}\left(
\begin{array}
[c]{c}%
v_{1}\\
u_{1}e^{-i\varphi_{1}}%
\end{array}
\right)  +\alpha_{0}\phi_{-q}\left(
\begin{array}
[c]{c}%
v_{2}\\
u_{2}e^{-i\varphi_{2}}%
\end{array}
\right)  \right]  ].\label{Eq6}%
\end{align}

Here $\varphi_{1,2}$ are the phases of the gaps $\Delta_{1,2}$ in both bands,
$u$ and $v$ are the standard Bogoliubov coefficients $u^{2}_{1,2}= \frac{1}%
{2}+\sqrt{E^{2}-\Delta_{1,2}^{2}}/2E,$ $v^{2}_{1,2}=\frac{1}{2}-\sqrt{
E^{2}-\Delta_{1,2}^{2}}/2E$. In the case of the $s_{\pm}$ gap model with
unequal s-wave gaps of opposite sign we have $\varphi_{1}-\varphi_{2}=\pi$,
while in the standard two-band model with the gaps of the same sign we have
$\varphi_{1}=\varphi_{2}.$ The amplitudes $a,b$ describe Andreev and normal
reflection, and the amplitudes $c,d$ describe transmission without branch
crossing and with branch crossing, respectively.

The total wave function must satisfy the following boundary conditions at the
interface ($x=0$)
\begin{align}
\Psi(0)=\Psi_{S}(0)  & =\Psi_{N}(0),\label{match1}\\
\frac{\hbar^{2}}{2m}\frac{d}{dx}\Psi_{S}(0)-\frac{\hbar^{2}}{2m}\frac{d}%
{dx}\Psi_{N}(0)  & =H\Psi(0),\label{match2}%
\end{align}
where $H$ is the strength of the (specular) barrier.

The boundary conditions on the wave function derivatives are usually expressed
in terms of Fermi velocities. However, a closer look reveals that this is
actually incorrect for Bloch waves. Therefore, in the following, we
introduce an \textquotedblleft interface velocity\textquotedblright. For a given Bloch function, say,
$\psi _{k}(x)=\sum_{G}A_{G,k}\exp [i(k+G)x]$, it is defined as
\begin{equation}
v_{k}=-\frac{i\hbar}{m}\frac{1}{\psi_{k}(x)}\frac{d\psi_{k}(x)}{dx}%
|_{x=0}.\label{velocity}%
\end{equation}
The so defined $v_{k}$ is real and has the same symmetry properties as the
Fermi velocity (this can be shown by expanding the wave functions in terms of
the plane waves), but it coincides with the actual group velocity only for
free electrons. For general Bloch waves it is different, and even dependent on
the  position of the interface plane in the crystal. We leave the interesting
and important issue of the relationship between the interface velocity and the
group velocity \cite{HF} for a further study, and proceed with the problem at hand.

Now, introducing the barrier strength $Z=H/\hbar v_{N}$, where $v_{N}$ is the
velocity on the N side, defined according to Eq. (\ref{velocity}), and using the
boundary conditions Eqs. (\ref{match1}, \ref{match2}), we find the general
solution for $a,b,c,$ and $d$. It depends on $Z$ and on the ratios of the
interface velocities. To keep the expressions compact, we present them below
for the case of equal interface velocities on the N side and in both bands on
the S side.

For the $s_{\pm}$ model where $\varphi_{1}-\varphi_{2}=\pi$ this gives
\begin{align}
\gamma a  & ={u_{1}v_{1}-\alpha(u_{1}v_{2}+u_{2}v_{1})+\alpha^{2}u_{2}v_{2}}%
{},\nonumber\\
\gamma b  & ={(Z^{2}+iZ)\left[  v_{1}^{2}-u_{1}^{2}+\alpha^{2}(u_{2}^{2}%
-v_{2}^{2})\right]  }{},\nonumber\\
\gamma c  & ={(1-iZ)(u_{1}-\alpha u_{2})}{}\delta,\nonumber\\
\gamma d  & ={iZ(v_{1}-\alpha v_{2})}{}\delta,
\end{align}
where $\gamma=(1+Z^{2})(u_{1}^{2}-\alpha^{2}u_{2}^{2})-Z^{2}(v_{1}^{2}%
-\alpha^{2}v_{2}^{2})$, $\delta=\psi_{k}(0)/\phi_{p}(0)$ and $\alpha
=\alpha_{0} \phi_{q}(0)/\phi_{p}(0).$ Note that for plane waves $\delta= 1$
and $\alpha= 0.$

For the $s_{++}$ model with $\varphi_{1}=\varphi_{2}$ we obtain
\begin{align}
\gamma a  & ={u_{1}v_{1}+\alpha(u_{1}v_{2}+u_{2}v_{1})+\alpha^{2}u_{2}v_{2}%
}{\ },\nonumber\\
\gamma b  & ={(Z^{2}+iZ)\left[  (v_{1}+\alpha v_{2})^{2}-(u_{1}+\alpha
u_{2})^{2}\right]  }{},\nonumber\\
\gamma c  & ={(1-iZ)(u_{1}+\alpha u_{2})}{\ }\delta,\nonumber\\
\gamma d  & ={iZ(v_{1}+\alpha v_{2})}{\ }\delta,
\end{align}
with $\gamma=(1+Z^{2})(u_{1}+\alpha u_{2})^{2}-Z^{2}(v_{1}+\alpha v_{2})^{2} $.

In a single band case ($\alpha=0$) and for plane waves the standard BTK
results are recovered. Below we shall discuss new effects arising in the
$s_{\pm}$ model. First, we consider a transparent interface, $Z=0$. In the
$s_{\pm}$\ case, $b=d=0,$ $a=(v_{1}-\alpha v_{2})/(u_{1}+\alpha u_{2}),$
$c=1/(u_{1}+\alpha u_{2})$. At $E=0$ we get $a=(\sqrt{\Delta_{1}}-\alpha
\sqrt{\Delta_{2}})/(\sqrt{\Delta_{1}}+\alpha\sqrt{\Delta_{2}})<1$, i.e.
Andreev reflection is suppressed. On the other hand, in the $s_{++}$ case
$b=d=0,$ $a=(v_{1}+\alpha v_{2})/(u_{1}+\alpha u_{2}),$ $c=1/(u_{1}+\alpha
u_{2})$ resulting in $a=1$ at zero energy, as expected in the standard BTK
model. This effect is due to the destructive interference between the
transmitted waves in the $s_{\pm}$ superconductor, which was missing in the
previous works \cite{Choi,Linder,Tsai,Feng,Ghaemi}.

If $\Psi=\left(
\begin{array}
[c]{c}%
f\\
g
\end{array}
\right)  $, then the probability current $J_{P}$ is given by
\begin{equation}
J_{P}=\frac{\hbar}{m}\left[  \text{Im}\left(  f^{\ast}\nabla f\right)
-\text{Im}\left(  g^{\ast}\nabla g\right)  \right]  ,
\end{equation}
properly taking electron and hole contributions into account. Using Eq.
(\ref{Eq6}) for $\Psi$ at the superconducting side of the interface,
$J_{P}=\left(  C+D\right)  J_{k}$,
where $J_{k}=v_{N}\left\vert \psi_{k}(0)\right\vert ^{2}$ is the probability
current of a normal electron in the state $\psi_{k}$, and the transmission
probabilities $C$ and $D$ depend on the velocities in the two bands. For equal
band velocities they are given by
\begin{align}
C &  =\left\vert c/\delta\right\vert ^{2}\left[  w_{1}+\alpha^{2}w_{2}%
+2\alpha\text{Re}\left(  u_{1}u_{2}^{\ast}\pm v_{1}v_{2}^{\ast}\right)
\right]  ,\\
D &  =\left\vert d/\delta\right\vert ^{2}\left[  w_{1}+\alpha^{2}w_{2}%
+2\alpha\text{Re}\left(  u_{1}u_{2}^{\ast}\pm v_{1}v_{2}^{\ast}\right)
\right]  ,
\end{align}
for the $s_{\pm}$ and $s_{++}$ models respectively, where $w_{1,2}=\left\vert
u_{1,2}\right\vert ^{2}-\left\vert v_{1,2}\right\vert ^{2}$.

At the normal side of the interface the probability current is $\left(
1-A-B\right)  J_{k}$, where $A=$ $\left\vert a\right\vert ^{2}$ and $B=$
$\left\vert b\right\vert ^{2}$ are Andreev and normal reflection
probabilities. From Eqs. (6,7) and (9,10), it can be verified that $A+B+C+D=1$.
Thus we have proven that the probability current is conserved. The electric
current $I$ across the contact is given by the standard expression \cite{BTK}
\begin{equation}
I=\frac{1}{eR_{N}}\int_{-\infty}^{\infty}{\left[  f_{0}\left(  E-eV\right)
-f_{0}\left(  E\right)  \right]  \left[  1+A-B\right]  dE},
\end{equation}
where $f_{0}$ is the Fermi function, $R_{N}$ is the normal state interface
resistance and $V$ the voltage bias across the interface. Below, we present
the results of calculations of the angle-resolved conductance $dI/dV$ in the
regime of a fully transparent interface $Z=0$ and in the tunneling regime
$Z\gg1$.

\begin{figure}
\includegraphics[width=7cm]{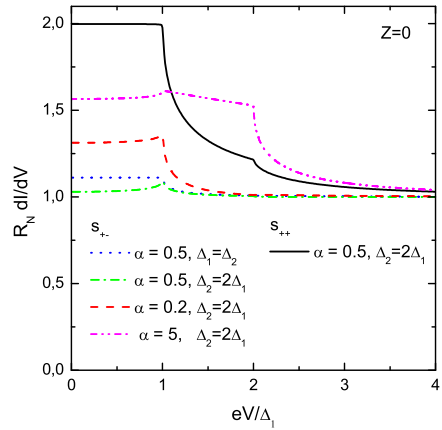}\caption{(Color online) Conductance in
the case of fully transparent interface, $Z=0$: comparison of the $s_{\pm}%
$\ and $s_{++}$ models.}%
\label{fig1}%
\end{figure}

The zero-temperature conductance at $Z=0$ is shown in Fig. 1. In the
$s_{++}$ case, there is a standard enhancement of conductance at low bias
$eV<\Delta_{1}$ due to Andreev reflection, followed by a decrease of
conductance and a weaker feature at $eV=\Delta_{2}$. At the same time, as
discussed above, a striking suppression of the zero-bias conductance occurs in
the $s_{\pm}$\ case. The strongest suppression occurs at $\alpha=\sqrt
{\Delta_{1}/\Delta_{2}}$.

\begin{figure}
\includegraphics[width=7cm]{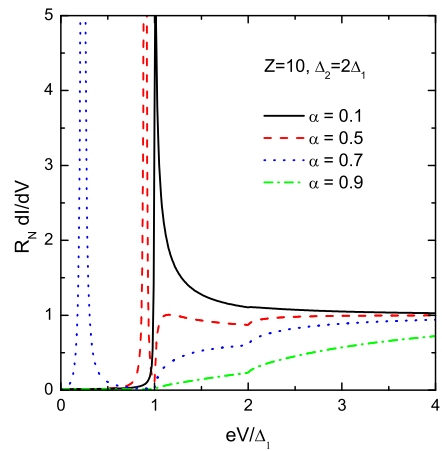}\caption{(Color online) Conductance in
the low transparency regime, $Z=10$, in the $s_{\pm}$\ model.}%
\label{fig2}%
\end{figure}

Figure 2 shows the zero-temperature conductance for large $Z$ in the $s_{\pm}$\ case.
Sharp conductance peaks appear at certain values of $\alpha$. These peaks have a clear interpretation as Andreev bound
states. Indeed, for large $Z$, a bound state exists if $\gamma=0$, that is,
if $u_{1}^{2}-v_{1}^{2}=\alpha^{2}(u_{2}^{2}-v_{2}^{2})$.
The energy of the bound state is
\begin{equation}
E_{B}=\sqrt{({\Delta_{1}^{2}-\alpha^{4}\Delta_{2}^{2}})/({1-\alpha^{4}})}.
\end{equation}
If $\Delta_{1}=\Delta_{2}$ this gives the trivial $E_{B}=\Delta$ solution,
that is, no subgap bound states. If $\alpha=0$, similarly, $E_{B}=\Delta_{1}$.
However, when $0\leq\alpha^{2}\leq\Delta_{1}/\Delta_{2}$ bound state solutions
exist (see Fig. 3), most notably a zero-bound state $E_{B}=0$ if $\alpha
^{2}=\Delta_{1}/\Delta_{2}$. Note the bound states for $\alpha=0.5$ and $0.7$
in Fig. 2.

\begin{figure}
\includegraphics[width=7cm]{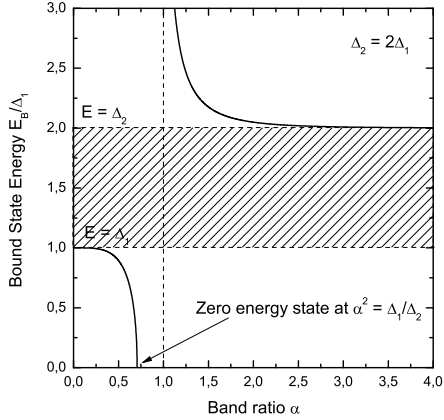}\caption{Bound state energy at contacts
to $s_{\pm}$ superconductors as function of the band ratio parameter $\alpha$
for $\Delta_{2}=2\Delta_{1}$. No bound states exist at energies between the
two gaps (shaded region). Also, for band ratios $\sqrt{\Delta_{1}/\Delta_{2}%
}<\alpha<1$ no surface bound states exist.}%
\label{fig3}%
\end{figure}

Further, it is also seen from Fig. 2 (\textit{e.g.,} for $\alpha=0.9$) that
there is a current deficit at high bias. This feature is due to a destructive
interband interference and is in contrast with the properties of N/S junctions
known so far, irrespective of whether S is s- or a d-wave. The only known case
is a double-barrier junction, where current deficit occurs due to
non-equilibrium quasiparticle distribution in the N layer \cite{SINIS}.

For comparison, we also present the results for the $s_{++}$ model in Fig. 4,
where bound states are absent, as expected. Still, interference effects at
$\alpha\sim1$ result in a complex $dI/dV$ behavior, where the conductance is
not a simple sum over two individual bands. Presently, the conductance spectra
of contacts with the multiband superconductor MgB$_{2}$ are usually fitted by
the sum of two single-band tunneling probabilities\cite{MgB2}. With the
increased level of sophistication in the realization of epitaxial magnesium
diboride junctions and single crystalline point contacts, one can expect that
the present predictions for multiband interference effects can be observed
there as well.
\begin{figure}
\includegraphics[width=6cm]{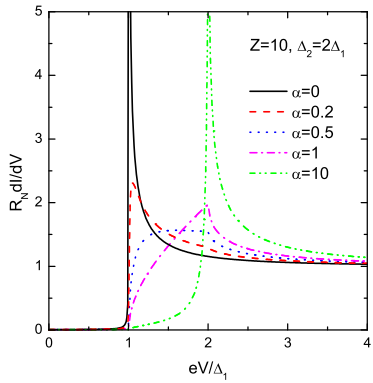}\caption{(Color online) Conductance in
the low transparency regime, $Z=10$, in the $s_{++}$ model.}%
\label{fig4}%
\end{figure}

In order to demonstrate the main features of the model, we have concentrated
on the discussion of the angle-resolved conductance. The total conductance
depends on the orientation of the interface and on the type of scattering,
specular or diffusive, which determines whether an electronic trajectory
crosses both bands or only one. Thus, knowledge of the junction geometry and
interface properties should make it possible to compare the model with
experimental data. Qualitatively, one can see already that the observation of
a zero-bias conductance peak can be consistent with nodeless
superconductivity, and that a non-zero energy surface bound state, that can
exist at subgap as well as supergap energies, can easily be mistaken for a gap
feature when interpreting conductance spectra.

This work was partially supported by the Netherlands Organization for Scientific Research
(NWO) and the NanoNed program under Grant No. TCS7029.

\end{document}